\documentclass[a4paper,fleqn,usenatbib]{mnras}

\usepackage{newtxtext,newtxmath}

\UseRawInputEncoding

\usepackage[T1]{fontenc}
\usepackage{ae,aecompl}
\usepackage{pdflscape}
\usepackage{rotating}
\usepackage{tablefootnote}
\usepackage{longtable}
\usepackage{threeparttable}
\usepackage{caption}
\usepackage{comment}

\usepackage{graphicx}	
\usepackage{amsmath}	
\usepackage{amssymb}	
\usepackage[LGRgreek]{mathastext}

\newcommand{\ch}{CH$_{3}$D}
\newcommand{\meth}{CH$_{4}$}

\title[\ch~detection in proto-BDs]{First \ch ~detection in Class 0/I proto-brown dwarfs: constraints on \meth~abundances}

\author[Riaz \& Thi]{
Riaz, B.$^{1}$\thanks{E-mail: briaz@usm.lmu.de}
\& Thi, W.-F.$^{2}$
\\
$^{1}$  Universit\"{a}ts-Sternwarte M\"{u}nchen, Ludwig Maximilians Universit\"{a}t, Scheinerstra$\beta$e 1, 81679 M\"{u}nchen, Germany
\\
$^{2}$  Max-Planck-Institut f\"{u}r Extraterrestrische Physik, Giessenbachstrasse 1, 85748 Garching, Germany
\\
}

\date{Accepted 2022 January 18. Received 2022 January 17; in original form 2021 November 11}

\pubyear{2021}

\begin{document}
\label{firstpage}
\pagerange{\pageref{firstpage}--\pageref{lastpage}}
\maketitle

\begin{abstract}

We report the first detection in the ${\it J}_{{\it K}}$ = 1$_{0}$ - 0$_{0}$ rotational transition line of \ch~ towards three Class 0/I proto-brown dwarfs (proto-BDs) from IRAM 30 m observations. Assuming a rotational temperature of 25 K, the \ch~abundances (relative to H$_{2}$) are in the range of (2.3--14.5) $\times$ 10$^{-7}$. The \meth~ abundances derived from the \ch~abundances and assuming the DCO$^+$/HCO$^+$ ratios are in the range of (0.05--4.8) $\times $10$^{-5}$. The gas-phase formation of \ch\ via CH$_2$D$^+$ is enhanced at high densities of 10$^{8}$--10$^{10}$ cm$^{-3}$ and our observations are likely probing the innermost dense and warm regions in proto-BDs. Thermal and/or non-thermal desorption can return the \ch~ and \meth~ molecules formed at an early stage on grain surfaces to the gas-phase. The gas phase abundances are indicative of warm carbon-chain chemistry in proto-BDs where carbon-chain molecules are synthesised in a lukewarm ($\sim$20-30 K) region close to the central source.

\end{abstract}


\section{Introduction}

Methane (\meth) has been fundamental in determining the properties of the coldest compact astrophysical sources, the T- and Y-type brown dwarfs (e.g., Cushing et al. 2011; Kirkpatrick et al. 2012). Their temperatures range between a few hundred kelvin and a few thousand kelvin, with most of their energy emitted at the infrared wavelengths. At these temperatures, brown dwarfs present strong methane absorption in their atmospheres, which is best probed by observing the \meth~rovibrational spectrum in the near-infrared, most notably the fundamental absorption bands near $\sim$1.58 $\mu$m and $\sim$3.3 $\mu$m. The presence and strength of near-infrared methane absorption in their atmospheres has been a key feature for the identification and classification of large numbers of Y- and T-type brown dwarfs.

Similar observations are difficult for early-stage Class 0/I proto-brown dwarfs (hereafter; proto-BDs) due to the high extinction and the excess continuum emission from the circumstellar material in the infrared that does not allow observing these photospheric \meth~ bands (e.g., Riaz \& Bally 2021). Also, \meth~does not have a pure rotational spectrum. The kinetic temperatures found in the cold, dense gas typical of molecular clouds and embedded Class 0/I cores are too low ($\sim$10-100 K) for the vibrational excitation of interstellar molecules, which makes it difficult to observe the infrared \meth~ bands under such conditions. Absorption study of near-IR \meth~ lines towards HL~Tau was unsuccessful with a 3 $\sigma$ upper limit \meth/CO ratio of 0.02\% (Gibb et al. 2004).

An alternative method to indirectly obtain methane abundance measurements is by observing its singly deuterated form, or mono-deuterated methane (\ch). \ch~ has a pure rotational spectrum and a non-zero small dipole moment (0.0057 Debye), which makes it possible to observe the \ch~rotational transitions from the ground (e.g., Womack et al. 1996). 

We report here the first detection in the ${\it J}_{{\it K}}$ = 1$_{0}$-0$_{0}$ rotational transition line of \ch~at millimeter wavelength of 232.6 GHz towards early-stage Class 0/I proto-BDs. In addition to \ch~abundance, we have obtained an estimate on the \meth~abundance by adopting the molecular DCO$^{+}$/HCO$^{+}$ ratios. It is interesting to investigate how much of the \meth~ seen in the field Y- and T-type brown dwarfs at a few hundred Myr to Gyr ages may have been inherited from the early formation stages. At the density and temperature of Y- and T-type brown dwarfs, it is most likely that the \meth~ observed in their atmosphere has been formed on site, i.e., completely reformed from C and H$_{2}$ after the accretion of the gas onto the central part of the proto-BD. Thus, there may not be a direct link between the \meth~ seen in proto-BDs and the one seen in older fully-formed brown dwarfs. 

Observations of \ch~ can also be used as a probe of the proto-BD environment, and to investigate whether \ch~ formed on grain surface and evaporated back to the gas-phase, or did it form in the warm $\sim$20-30 K regions relatively close to the central object (e.g., Womack et al. 1996; Sakai et al. 2012; Roueff et al. 2013; Asvany et al. 2004; Cleeves et al. 2016; Qasim et al. 2020; G\"{a}rtner et al. 2010). 

Our IRAM 30 m observations are presented in Sect.~\ref{obs} and the results thus obtained are presented in Sect.~\ref{results}. Section~\ref{reactions} presents the formation pathway for the \ch~ and \meth~ molecules in the gas- and grain-phase, and Section~\ref{discuss} investigates the possibility of the Warm Carbon Chain Chemistry (WCCC; Sakai et al. 2008) scenario in these cool, dense objects.

\section{Targets, Observations, and Data Reduction}
\label{obs}

We observed the ${\it J}_{{\it K}}$ = 1$_{0}$-0$_{0}$ line (232.644301 GHz, Womack et al. 1996) of \ch~ towards the complete sample of 16 Class 0/I proto-BDs described in Riaz \& Thi (2022). The observations were obtained at the IRAM 30 m telescope in December, 2017, and February, 2019. We used the EMIR heterodyne receiver (E230 band), and the FTS backend in the wide mode, with a spectral resolution of 200 kHz ($\sim$0.3 km s$^{-1}$ at 232 GHz). The observations were taken in the frequency switching mode with a frequency throw of approximately 7 MHz. The source integration times ranged from 3 to 4 hours per source per tuning reaching a typical RMS (on {\it T}$_{A}^{*}$ scale) of $\sim$0.01-0.02 K. The telescope absolute RMS pointing accuracy is better than 3$\arcsec$ (Greve et al. 1996). All observations were taken under good weather conditions (0.08 $<$ $\tau$ $<$0.12; PWV $<$2.5 mm). The absolute calibration accuracy for the EMIR receiver is around 10\% (Carter et al. 2012). The telescope intensity scale was converted into the main beam brightness temperature ({\it T}$_{mb}$) using standard beam efficiency of $\sim$59\% at 230 GHz. The half power beam width of the telescope beam is $\sim$10$\arcsec$ at 230 GHz. The spectral reduction was conducted using the CLASS software (Hily-Blant et al. 2005) of the GILDAS facility\footnote{http://www.iram.fr/IRAMFR/GILDAS}. The standard data reduction process consisted of averaging multiple observations, extracting a subset around the line rest frequency, and fitting a low-order (4-6) polynomial baseline which was then subtracted from the average spectrum. We chose a range of $\pm$10 km/s around the source {\it V}$_{lsr}$ to fit the baseline.

 \begin{figure*}
  \centering   
     \includegraphics[width=2.2in]{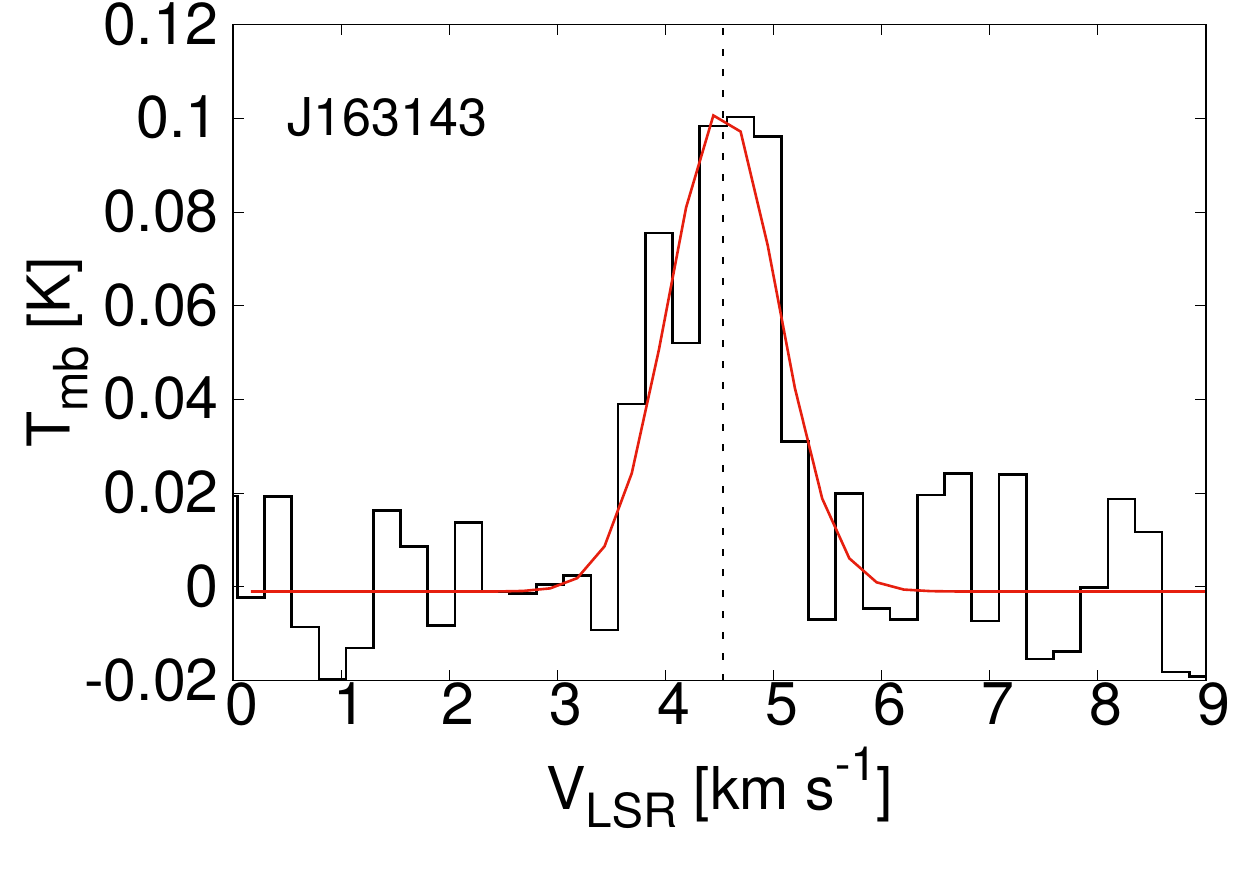}
     \includegraphics[width=2.2in]{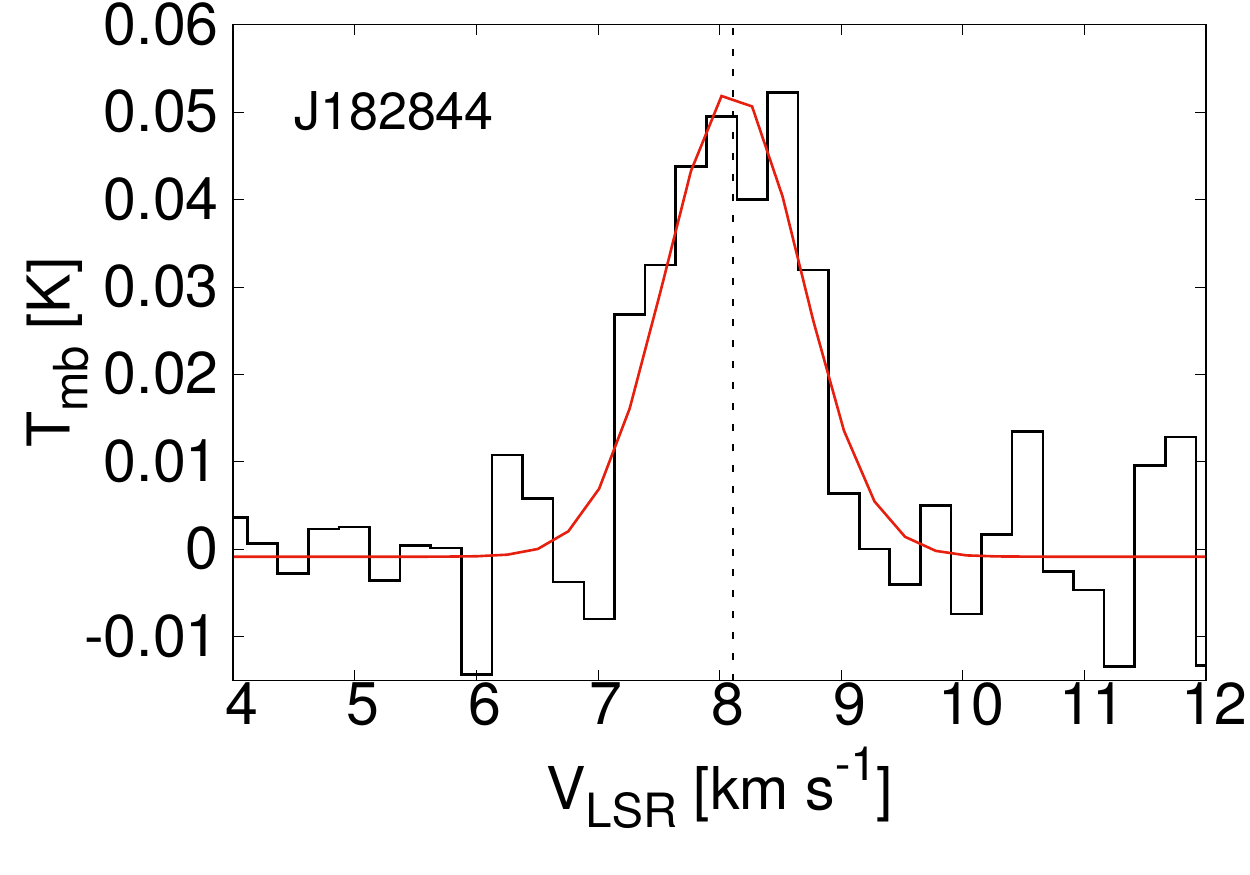}
     \includegraphics[width=2.2in]{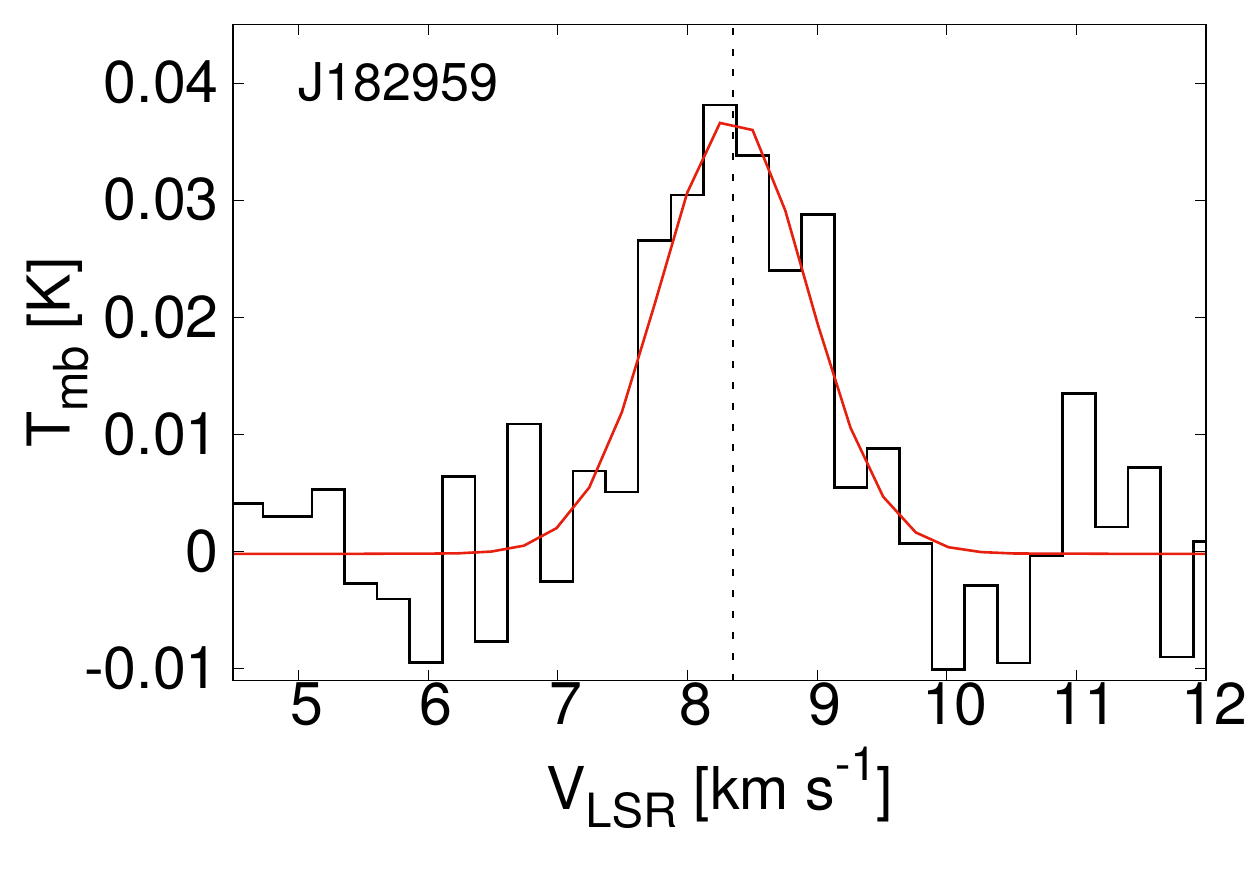}          
     \caption{The observed CH$_{3}$D spectra (black) with the Gaussian fits (red). Black dashed line marks the cloud systemic velocity of $\sim$8 km/s in Serpens and $\sim$4.4 km/s in Ophiuchus. }
     \label{spectra}     
  \end{figure*}

At present, reaching the same level of sensitivity of $\sim$10-20 mK, there are only three proto-BDs in our sample of 16 objects that show a $\geq$3-$\sigma$ detection in \ch: SSTc2d J163143.8-245525 (J163143; {\it L}$_{bol}\sim$0.08 {\it L}$_{\sun}$), SSTc2d J182844.8+005126 (J182844; {\it L}$_{bol}\sim$0.04 {\it L}$_{\sun}$), and SSTc2d J182959.4+011041 (J182959; {\it L}$_{bol}\sim$0.008 {\it L}$_{\sun}$). A detailed discussion on the classification and mass and luminosity measurements for these objects can be found in Riaz et al. (2018) and Riaz \& Thi (2022).

\section{Results}
\label{results}

Figure~\ref{spectra} shows the \ch~spectra for the three proto-BDs. We have measured the parameters of the line center, line width, the peak and integrated intensities using a single-peaked Gaussian fit. The apparent double-peak shape could be due to the noise in the spectrum. The line parameters are listed in Table~\ref{pars}. The uncertainty is estimated to be $\sim$20\% for the peak and integrated intensities and $\Delta${\it v}, and $\sim$0.02-0.04 km/s for {\it V}$_{lsr}$. The errors on the line parameters are due to uncertainties in fitting the line profile, and mainly arise from the end points chosen for the Gaussian fit.

As a check, we have compared the \ch~ {\it V}$_{lsr}$ with those measured from the C$^{17}$O (2-1) observations (Riaz et al. 2019). The C$^{17}$O line is optically thin and provides a reliable reference for the source {\it V}$_{lsr}$. The C$^{17}$O {\it V}$_{lsr}$ is 4.6$\pm$0.02 km/s for J163143, 7.8$\pm$0.02 km/s for J182844, and 8.4$\pm$0.02 km/s for J182959. The C$^{17}$O {\it V}$_{lsr}$ for J163143 and J182959 are consistent with those measured for the \ch~ lines, while a shift of $\sim$0.3 km/s is seen for J182844. This shift is likely due to the broad, non-Gaussian shape of the \ch~ profile for J182844 compared to the Gaussian-like C$^{17}$O line. Note that the \ch~ FWHM is twice that of C$^{17}$O for J182844, and the shift in {\it V}$_{lsr}$ is within the FWHM. 

We had also obtained in the same observing runs spectra at four offset positions around these targets. The off-source spectra were taken at a step size of 1-2 beamsize ($\sim$10$\arcsec$) offset from the target position. There is no detection in the CH$_{3}$D line at a $\geq$2-$\sigma$ level (1-$\sigma$ rms $\sim$10-20 mK) at any offset position. We can therefore be confident that the observed CH$_{3}$D emission has an origin from the source position. There could still be foreground/background contamination within the beamsize in the on-source spectrum but we cannot quantify it at present without interferometric observations.

The column density of \ch~ is derived from the integrated line intensity by assuming optically thin emission, local thermodynamic equilibrium (LTE) condition, and excitation temperatures of 10 K, 15 K, 20 K, and 25 K. The method for deriving the \ch~column density is described in the Appendix. The kinetic temperature is expected to be relatively low ($\sim$10 K) throughout the outer and inner envelope layers in the proto-BDs (e.g., Machida et al. 2009) with only the innermost ($<$20 au) region to be at higher temperatures ($>$ 20~K).

The \ch~column densities and the molecular abundances relative to H$_{2}$, [{\it N}(X)/{\it N}(H$_{2}$)] are listed in Table~\ref{abund}. The H$_{2}$ column densities are computed using the source size and the (dust + gas) mass derived from the sub-millimeter continuum emission (Riaz et al. 2018). The \ch~ and \meth~ abundances for the proto-BDs are higher by a small factor of $\sim$1.6 for an excitation temperature of 25 K compared to 10 K (Table~\ref{abund}). Also listed in Table~\ref{abund} are the \meth~ abundances that were derived from [\ch] and the DCO$^{+}$/HCO$^{+}$ ratios for the proto-BDs (Riaz \& Thi 2022). The error on the column densities is propagated from the error on the integrated line intensity, and is estimated to be $\sim$20\%. The error on the \ch~ abundances is propagated from the error on the \ch~ and H$_{2}$ column densities, and is estimated to be $\sim$23\%-24\%. The error on the \meth~ abundances is propagated from the error on the \ch~ abundance and the D/H ratio, and is estimated to be $\sim$30\%-32\%.

\begin{table*}
\centering
\caption{CH$_{3}$D Line Parameters \tnote{a}}
\label{pars}
\begin{threeparttable}
\begin{tabular}{llllllll} 
\hline
Object   	& {\it V}$_{lsr}$ 		& {\it T}$_{mb}$		& $\int{{\it T}_{mb} dv}$ 	&  $\Delta${\it v}  \\
	     	& (km s$^{-1}$)		& (K)				& (K km s$^{-1}$) 	& (km s$^{-1}$)		\\
\hline
J163143	& 4.53$\pm$0.02	& 0.13$\pm$0.02	& 0.13$\pm$0.02	& 1.01$\pm$0.24		\\
J182844	& 8.11$\pm$0.03	& 0.07$\pm$0.01	& 0.08$\pm$0.01	& 1.12$\pm$0.25		\\
J182959	& 8.35$\pm$0.02	& 0.05$\pm$0.01	& 0.06$\pm$0.01	& 1.14$\pm$0.25		\\
\hline
\end{tabular}
\begin{tablenotes}
\item[a] The line parameters have been derived from a single-gaussian fit to the profiles.
\end{tablenotes}
\end{threeparttable}
\end{table*}

\begin{table*}
\centering
\caption{Column Densities and Molecular Abundances}
\label{abund}
\begin{threeparttable}
\begin{tabular}{ccc|cccc|cccc|cccc} 
\hline
Object & ${\it N}_{H_{2}}$  & D/H \tnote{a} & \multicolumn{4}{c}{${\it N}$(CH$_{3}$D) (x10$^{16}$ cm$^{-2}$) \tnote{b}} 	& \multicolumn{4}{c}{[CH$_{3}$D] (x10$^{-7}$) \tnote{b}} 	& \multicolumn{4}{c}{[CH$_{4}$] (x10$^{-5}$) \tnote{b}}		\\
\hline
	   & (x10$^{22}$ cm$^{-2}$) & & 10 K & 15 K & 20 K & 25 K & 10 K & 15 K & 20 K & 25 K & 10 K & 15 K & 20 K & 25 K	\\
\hline
J163143	& 1.8$\pm$0.2 	& 0.03$\pm$0.006 & 1.6 & 1.8 & 2.1 & 2.6 & 8.7 & 9.9 & 12.2 & 14.5 & 2.9 & 3.3 & 4.0 & 4.8	\\
J182844	& 3.6$\pm$0.5 	& 0.6$\pm$0.1 & 0.8 & 1.0 & 1.2 & 1.4 & 2.4 & 2.8 & 3.3 & 4.0 & 0.04 & 0.04 & 0.05 & 0.06	\\
J182959	& 4.5$\pm$0.5	& 0.4$\pm$0.09 & 0.6 & 0.7 & 0.9 & 1.0 & 1.4 & 1.7 & 1.9 & 2.3 & 0.03 & 0.04 & 0.04 & 0.05	\\ 			
\hline
\end{tabular}
\begin{tablenotes}
\item[a] The [DCO$^{+}$/HCO$^{+}$] ratios are from Riaz \& Thi (2022). 
\item[b] The errors are estimated to be $\sim$20\% on the column densities, $\sim$23\%-24\% on the \ch~ abundances, and $\sim$30\%-32\% on the \meth~ abundances. 
\end{tablenotes}
\end{threeparttable}
\end{table*}

\section{CH$_4$ and CH$_3$D formation pathways}
\label{reactions}

\begin{figure*}
  \centering   
     \includegraphics[width=4.5in]{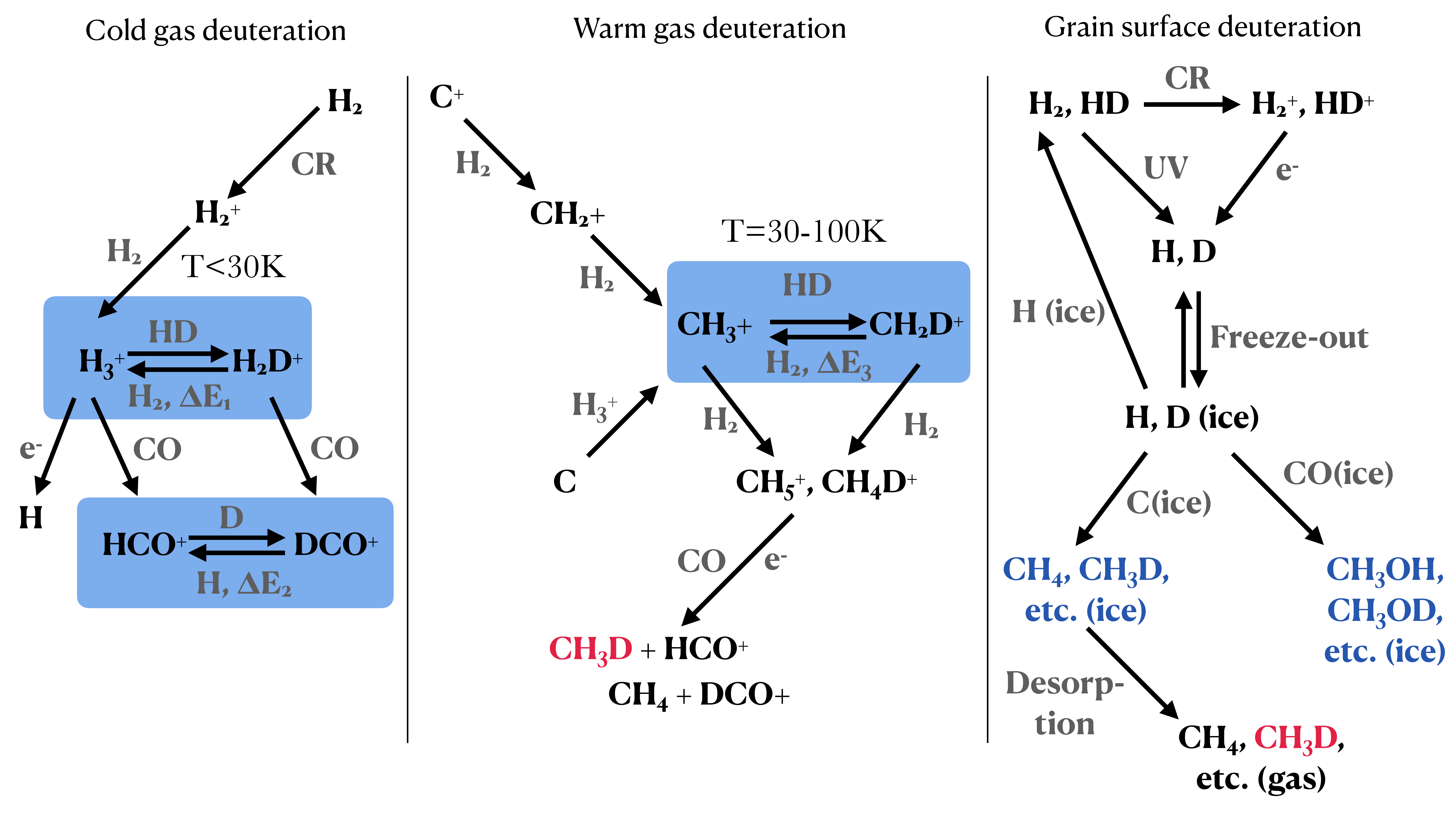}
     \caption{A diagram of the formation pathways for \ch~ and \meth~ via gas-phase and grain surface reactions. $\Delta$ E1, E2, E3 are positive (i.e. the reactions are exothermic). Diagram is based on the reactions proposed by Roueff et al. (2013), Albertsson et al. (2013), Cleeves et al. (2016), Asvany et al. (2004), and Adams \& Smith (1985). }
     \label{network}     
  \end{figure*}

Figure~\ref{network} shows the potential gas and surface formation pathways for \meth\ and \ch. The cold gas deuteration reactions via H$_2$D$^+$ are discussed in Riaz \& Thi (2022). The gas-phase formation pathway for \ch~ and \meth~is initiated by the reaction between C$^+$ and H$_2$ (e.g., Roueff et al. 2013; Cleeves et al. 2016; Asvany et al. 2004):
\begin{gather}
C^+ + H_{2} \rightarrow CH_2^+\\
CH_2^{+} + H_{2} \rightarrow CH_3^{+} + H
\end{gather}
\noindent CH$_3^+$ reacts with H$_2$ via the slow radiative association reaction:
\begin{gather}
CH_3^{+} + H_2 \rightarrow CH_5^{+} + h\nu
\end{gather}
\noindent or can exchange a deuteron atom with HD:
\begin{gather}
CH_3^{+} + HD \leftrightarrows CH_2D^{+} + H_2 + \Delta	E_1
\end{gather}
\noindent The rate coefficient of reaction (3) decreases with increasing temperature (KIDA database: Wakelam et al. 2012, http://kida.astrophy.u-bordeaux.fr). Since the forward reaction (4) is exothermic (with $\delta E_1$ between 483 and 660~K depending on the ortho-para ratio of $CH_3^{+}$ and H$_2$, Roueff et al. 2013), it is in general favoured in the temperature range up to 80~K (Albertsson et al. 2013). However, because the exothermicity is lower for ortho-H$_2$, the reverse reaction becomes possible at lower temperatures when H$_2$ is mostly in the ortho state. Similar to CH$_3^+$,
\begin{gather}
CH_2D^{+} + H_2 \rightarrow CH_4D^+ + h\nu
\end{gather}
\noindent the recombination reaction with free electrons
\begin{gather}
CH_5^+ + e^- \rightarrow CH_4\\
CH_4D^+ + e^- \rightarrow CH_3D
\end{gather}
\noindent or on negatively-charged grain surfaces
\begin{gather}
CH_5^+ + grain^- \rightarrow CH_4\\
CH_4D^+ + grain^- \rightarrow CH_3D
\end{gather}
\noindent results in \meth\ and \ch. Reaction with CO can lead to HCO$^+$ and DCO$^+$:
\begin{gather}
CH_5^+ + CO \rightarrow HCO^+ + CH_4\\
CH_4D^+ + CO \rightarrow DCO^+ + CH_4
\end{gather}

Apart from the two radiative association reactions, the reactions involve an ion and a neutral species and proceed at Langevin rate. The rates are of the order of 10$^{-10}$--10$^{-9}$ cm$^{3}$/s and are temperature-independent. If the grain temperature is below the desorption temperature of 
$\sim$25 K (Collings et al. 2004), CH$_4$/\ch~ will freeze-out with a typical timescale of $\sim$3000 yrs at 10~K and density of 10$^{6}$ cm$^{-3}$.

Alternatively, CH$_4$ and CH$_3$D can be desorbed from CH$_4$(ice) and CH$_3$D(ice) formed on grain surfaces. The formation of \meth~ on ice grains is via successive H or D additions of carbon on grain surfaces (e.g., Qasim et al. 2020). At each stage, it is a competition between attaching H(ice) or D(ice). Once formed, CH4(ice) does not react with HD or D atom to form CH3D(ice) due to the endothermicity of the substitution reaction or the existence of barrier (e.g. Li et al. 2015), respectively.

\section{Discussion}
\label{discuss}

The main gas-phase channels for the formation of \ch~ and \meth~ is initiated by the formation of CH$_{5}^{+}$ and CH$_{4}$D$^{+}$ via the radiative association reactions (3) and (5), which are enhanced at high densities of 10$^{8}$--10$^{10}$ cm$^{-3}$ (Cleeves et al. 2016), because the competing election recombination reactions are hampered by low electron fractions. The ionization rate is proportional to the gas density, while the rate of recombination with electrons depends on the square of the density making the electron density at high densities very low. The reactions are more efficient at low and moderate temperatures ({\it T} $< \sim$~40~K).

The detection of \ch~ suggests therefore that we may be probing the innermost densest regions of a proto-BD, close to the jet launching points ($<$20 au), where the densities are $\geq$10$^{8}$ cm$^{-3}$ and the temperature can reach values of $\sim$20-30 K (Machida et al. 2009). In addition to the gas-phase formation, thermal (when the dust temperature is above 25~K) and/or non-thermal desorption can also return the \ch~ and \meth~ molecules formed on grain surfaces at an earlier evolutionary stage to the gas phase. The abundances for proto-BDs in Table~\ref{abund} are comparable to the \ch~($\sim$3$\times$10$^{-7}$) and \meth~($\sim$4$\times$10$^{-6}$) abundances reported by Sakai et al. (2012) for the low-mass protostar IRAS04368+2557 in L1527.  Therefore a Warm Carbon Chain Chemistry may occur in proto-BDs with carbon-chain molecule formation in a lukewarm ($\sim$20-30 K) region close to the central proto-BD, similar to the low-mass protostars (Sakai et al. 2008; 2009; 2012).

The proto-BD J163143 (ISO-OPH 200) has been modelled in details by Riaz \& Machida (2021). Non-thermal desorption is most likely at play for J163143, which drives a strong jet/outflow. This proto-BD shows a factor of 3-4 higher \ch~abundance and about 2 orders of magnitude higher \meth~abundance than the other two proto-BDs. Interestingly, this proto-BD also has a D/H ratio that is about an order of magnitude lower than the other two proto-BDs (Table~\ref{abund} and Riaz \& Thi 2022). There may be a possible dependency between the WCCC and a low deuterium fractionation, as already noted for the case of the low-mass protostar IRAS04368+2557 (Sakai et al. 2012). ISO-OPH 200 is a proto-BD with a very young kinematic age of $\sim$6000 yr, and as discussed in Riaz \& Machida (2021), the embedded lifetime appears shorter for proto-BDs compared to low-mass protostars, with an earlier transition from the Class 0/I to the Class II phase at $<$0.05 Myr rather than $\sim$0.1-0.5 Myr as typically seen for low-mass protostars. A short embedded phase will lower the formation efficiency of molecules on grain surfaces. If indeed we have a case of fast core contraction, then carbon atoms depleted onto dust grains will react with H via grain surface reactions to form \meth(ice) before they are converted to CO in the gas phase (e.g., Sakai et al. 2008; 2009). A short embedded life would also imply that the timescale for core contraction is not long enough for deuterium transfer reactions to occur. Together, these scenarios could explain the high \ch~ and \meth~ abundance and low D/H ratio for ISO-OPH 200. Another point to note is the lack of CCS and HC$_{3}$N detection in these proto-BDs (Riaz et al. 2018; 2019), which is also consistent with the WCCC scenario (e.g., Sakai et al. 2008).

The lack of detection in \ch~ at a $>$2-$\sigma$ level for the remaining 13/16 proto-BDs can either be due to the poor line sensitivity of the observations or the absence of a warm source close to the central proto-BD, such as a high velocity jet, which can warm up the central region and result in the photoevaporation or cosmic-ray-induced evaporation of \meth~ and trigger WCCC. Most of methane and CH$_3$D molecules in these proto-BDs may still be in the ice form. 

Observations of the ${\it J}_{{\it K}}$ = 2$_{0}$ - 1$_{0}$ and ${\it J}_{{\it K}}$ = 2$_{1}$ - 1$_{1}$ lines at 465 GHz will secure the detection and confirm that the emitting gas is warm. Detection of carbon-chain molecules in the proto-BD with detected CH$_3$D will test that WCCC occurs in the innermost region of proto-BDs. Finally, sensitive interferometric observations will help to understand the spatial distribution of these molecules and their origin of emission. 


\section*{Acknowledgements}

We thank the anonymous referee for several insightful comments on the paper. B.R. acknowledges funding from the Deutsche Forschungsgemeinschaft (DFG) - Projekt number RI-2919/2-1. This work is based on observations carried out with the IRAM 30 m telescope. IRAM is supported by INSU/CNRS (France), MPG (Germany) and IGN (Spain). 

\section{Data Availability}

The data underlying this article are available in the IRAM archives through the VizieR online database.

\section{Appendix}

Assuming optically emission, the column density of CH$_3$D in cm$^{-2}$ without correcting for beam dilution effects, is calculated from the line flux of the transition ${\it J}_{\it K} \rightarrow ({\it J}-1)_{\it K}$ 
with ${\it J}=1$ and  ${\it K}=0$ using the formula (Womack et al. 1996):
\begin{equation}
{\it N}({\rm CH_3D}) = \frac{3\times10^5 k}{8 \pi^3 \nu \mu_0^2}\frac{{\it J}}{{\it J}^2-{\it K}^2}\frac{Q_{\rm rot(T_{\rm ex})} e^{h\nu/kT_{\rm ex}}}{e^{\Delta E/kT_{\rm rot}}S_{I, K}}\int {\it T}_{\rm mb}\ dv,
\end{equation}
where $\nu$ is the frequency of the transition in Hz (232.6443010 GHz, Womack et al. 1996), $\mu_0$ is the dipole moment in Statcoulomb-centimeters (1 Debye = 10$^{-18}$ StatC cm), $k$ is the Boltzmann constant (cgs), $h$ is the Plank constant (cgs), ${\it J}$ and ${\it K}$ are the quantum numbers, ${\it T}_{\rm ex}$ is the excitation temperature, ${\it T}_{\rm rot}$ is the excitation temperature, $Q_{\rm rot(T_{\rm ex})}$ is the partition function at ${\it T}_{\rm ex}$, $\Delta E$ is the energy in ergs of the ${\it J}-1$ lower level with respect to the ground level, $S_{I, K}$ is the statistical weight factor. ${\it T}_{\rm mb}$ is the main-beam temperature in Kelvin and $\int {\it T}_{\rm mb}\ dv$ is the line flux in K km s$^{-1}$. Similar to (Womack et al. 1996), the microwave background was neglected and we assumed ${\it T}_{\rm ex}={\it T}_{\rm rot}$.

The statistical weight factor is caused by the presence of three equivalent nuclei with $I = 1/2$ (Townes \& Schawlow 1975; Womack et al. 1996):
\begin{equation}
S_{I, K} = \frac{2(4I^2+4I)}{4I^2+4I+1}
\end{equation}
The line frequencies, energy levels, and the dipole moment ($\mu_0$=0.0059 Debye) are taken from the CDMS database (Endres et al. 2016). For the ${\it J}=1$, ${\it K}=0$ transition, $\Delta E = 0$ ergs. The value at ${\it T}_{\rm ex}$ of the partition function was calculated using a log-log interpolation of the value in the CDMS database (Endres et al. 2016) instead of an approximate formula. With those values, we reproduced the upper limit from the ${\it J}=2$, ${\it K}=1$ transition in Womack et al. (1996) within 30\%.


\bsp	
\label{lastpage}
\end{document}